\documentclass[prl,lengthcheck,onecolumn,superscriptaddress]{revtex4-2}
\usepackage[T1]{fontenc}
\usepackage[utf8]{inputenc}
\usepackage[a4paper,margin=20mm]{geometry}
\usepackage{amsmath,amssymb,bm,mathtools}
\usepackage{graphicx,booktabs,array,tabularx,placeins,float}
\usepackage{microtype,setspace,caption,titlesec}
\usepackage{times}
\graphicspath{{figures/}}
\titleformat{\section}{\large\bfseries}{}{0pt}{}
\titleformat{\subsection}{\normalsize\bfseries}{}{0pt}{}
\titlespacing*{\section}{0pt}{1.3em}{0.5em}
\titlespacing*{\subsection}{0pt}{0.8em}{0.2em}
\newcommand{\ImPart}{\mathrm{Im}}
\newcommand{\ii}{\mathrm{i}}

\newcommand{\G}{\mathbf{G}}
\newcommand{\Gz}{\mathbf{G}_0}
\newcommand{\Vso}{\mathbf{V}_{\mathrm{so}}}
\newcommand{\Hzero}{\hat{H}_0}
\newcommand{\Htot}{\hat{H}}

\newcommand{\ket}[1]{\left|#1\right\rangle}

\newcommand{\mel}[3]{\left\langle#1\middle|#2\middle|#3\right\rangle}
\DeclareMathOperator{\Tr}{Tr}
\newcommand{\compound}{Sr$_3$Co$_2$O$_7$}
\newcommand{\papertitle}{Giant magnetic anisotropy in layered polar metal Sr$_3$Co$_2$O$_7$  from Charge-disproportionation orbital unquenching}
\usepackage{xr-hyper}

\usepackage[colorlinks=true,citecolor=blue,linkcolor=blue,urlcolor=blue]{hyperref}

\usepackage{cleveref}
\begin{document}
\begin{center}
{\LARGE\bfseries \papertitle\par}
\vspace{0.8em}
Houssam Sabri$^{1,2,*}$, Hong-Fei Huang$^3$, Jie-Xiang Yu$^{3}$ and Jiadong Zang$^{1}$
\end{center}
{\small
$^1$Department of Physics and Astronomy, University of New Hampshire, Durham, New Hampshire 03824, USA.\\
$^2$College of Physical Sciences and Engineering, Mohammed VI Polytechnic University, Benguerir, 43150, Morocco.\\
$^3$School of Physical Science and Technology, Soochow University, Suzhou 215006, China.\\
$^*$Correspondence: houssam.sabri@um6p.ma.
}

\section{Abstract}
Crystal-field quenching usually suppresses orbital angular momentum and limits magnetic anisotropy in transition-metal oxides. Here we identify charge disproportionation as a route to giant magnetocrystalline anisotropy in the layered polar metal \compound. First-principles calculations reveal an energy preference of 234.0 meV per cell for in-plane [110] magnetization over [001]. Charge disproportionation and the associated polar displacement reconstruct the local cobalt crystal fields, creating inequivalent sites with distinct orbital hierarchies. This reconstruction enables a strongly directional orbital response, particularly at the displaced cobalt site, where orbital angular momentum is recovered for in-plane magnetization. Spin--orbit coupling converts this response into a large magnetic anisotropic energy. In the centrosymmetric reference, the orbital moments remain nearly quenched and the anisotropy is strongly reduced. A local Green-function and orbital-moment analysis connects the large energy scale to orbital unquenching, while the conventional second-order contribution is small within the adopted approximation. These results identify charge-disproportionation-driven orbital reconstruction as a design principle for large anisotropy and motivate layered perovskites and magnetic polar metals as platforms in the search for rare-earth-free permanent magnets.

\section{Introduction}
Polar metals challenge the conventional separation between structural polarity and itinerant charge. Whereas metallic carriers screen static electric fields, they do not necessarily prevent a crystal from adopting a polar, inversion-breaking structure \cite{devonshire_theory_1954,anderson_symmetry_1965}. The coexistence of polarity and conductivity creates an opportunity to couple structural symmetry breaking to electronic and magnetic degrees of freedom. A central question is how this coupling can generate a large functional response, particularly when the same correlated electrons participate in both the structural reconstruction and magnetism.

Layered perovskites provide a natural setting for exploring this question. Their perovskite slabs alternate with chemically distinct spacer layers, producing an intrinsically anisotropic architecture. Across the Ruddlesden--Popper, Dion--Jacobson, Aurivillius and Carpy--Galy families, structural flexibility supports diverse polar states, while advances in epitaxial growth open opportunities to investigate polar metallicity and coupled ferroic responses \cite{gradauskaite_layered_2026}. In magnetic members of these families, changes in local coordination and bonding can reshape the transition-metal crystal fields and the orbital degrees of freedom that couple magnetization to the lattice.

Magnetocrystalline anisotropy energy (MAE) directly probes this coupling. It sets an intrinsic energy scale for magnetic orientation and is important for magnetic storage and spintronic devices \cite{chappert_emergence_2007,brataas_current_2012,dieny_perpendicular_2017}. Large anisotropy is also an essential ingredient in the search for rare-earth-free permanent magnets. In many $3d$ magnets, however, MAE is much smaller than the atomic spin--orbit coupling (SOC) scale because the crystal field quenches orbital angular momentum. The leading orientation dependence then arises through virtual transitions between occupied and unoccupied states and is quadratic in SOC strength \cite{wang_first_principles_1993,bruno_tight_binding_1989,van_der_laan_microscopic_1998,Li2015}. An orbitally active manifold offers a different regime: SOC can act within near-degenerate states and generate a large directional orbital response. The resulting energy depends on occupation, crystal-field splitting and bandwidth; in appropriate atomic or flat-band limits, a first-order contribution can survive \cite{huang_magnetic_2022}. Recent first-principles studies have demonstrated that the combined action of electronic correlations and self-consistent SOC can stabilize large unquenched orbital moments and enhance magnetic anisotropy in layered transition-metal compounds \cite{Ovesen2024}. The emergence of these moments depends on crystal-field symmetry and orbital filling, with partially occupied $t_{2g}$ manifolds supporting complex orbital combinations and first-order SOC splitting \cite{Ovesen2026}. Experimentally, X-ray magnetic dichroism measurements on VI$_3$ have revealed a substantial unquenched orbital moment and its evolution near a spin-reorientation transition, highlighting the connection between orbital degrees of freedom and magnetic orientation \cite{DeVita2024}.

Several routes to polar metallicity separate the structural instability from the states responsible for conduction, for example through orbital or geometric design \cite{kim_polar_2016,wetherington_2-dimensional_2021,kolodiazhnyi_persistence_2010}. Related studies have explored photoinduced metal--insulator transitions, strain-gradient control, correlated ferromagnetic polar states, and optical identification of noncentrosymmetric domains\cite{deng_polarization-dependent_2024,zabalo_switching_2021,zhang_correlated_2024,tang_structural_2020}. These developments motivate a complementary question: can an electronic reconstruction that supports polarity also activate a large magnetic anisotropy?

The broader importance of magnetic orientation is illustrated by skyrmion crystals\cite{muhlbauer_skyrmion_2009,yu_real-space_2010}, and topological Hall responses \cite{neubauer_topological_2009,yin_topological_2015,vistoli_giant_2018,shao_topological_2019}, whose behavior is sensitive to magnetic structure. Thermal fluctuations, frustration and defects further influence magnetic textures \cite{hou_thermally_2017,yu_thermally_2019,SabriPRL,SabriPRB,Channa2025}, while sliding ferroelectrics and layered orbital order connect structural or orbital organization to electronic topology and spin splitting\cite{qian_switchable_2025,Meier2026}. Together, these studies motivate control of the microscopic interactions that select magnetic orientation.

Charge disproportionation provides a candidate link between structural reconstruction and anisotropy. It creates electronically inequivalent transition-metal sites, modifies local bonding and changes crystal-field splittings and spin-dependent occupations. Its connection to structural and metal--insulator transitions is well established in correlated oxides \cite{takano_charge_1977,woodward_structural_2000,lee_disproportionation_2005,ZHANG2024124865,torrance_systematic_1992,alonso_charge_1999,mazin_charge_2007}. In a magnetic polar metal, the same redistribution can reconstruct the orbital states on which SOC acts. The relevant design objective is therefore a charge-disproportionated environment that retains an orbitally responsive manifold near the Fermi level.

The double-layered Ruddlesden--Popper cobaltate \compound\ provides a concrete platform \cite{zhou_geometry-driven_2025}. First-principles work identifies charge disproportionation and an inverted local $t_{2g}$ hierarchy as a route to polar metallicity \cite{ChargeDiso}. and related charge disproportionation mechanisms have been considered in other polar magnetic metals \cite{shen2026}. Here we investigate the ferromagnetic electronic configurations of \compound\ and show that the charge-disproportionated polar state supports giant anisotropy accompanied by a large, site-selective orbital response. Comparison with a centrosymmetric reference and local SOC analysis connect this enhancement to orbital unquenching. The resulting picture identifies a microscopic route by which layered perovskites and polar metals can contribute to the search for strongly anisotropic magnets without rare-earth elements.

\section{Results}
\subsection{Charge disproportionation reconstructs the local orbital environment}
The polar phase (POL) contains two inequivalent cobalt sublattices, Co1 and Co2, with two sites of each type in the conventional cell. Co1 is displaced along the stacking direction and has an asymmetric apical oxygen environment. The corresponding change in Co--O bonding produces a local crystal field distinct from that at Co2. This site differentiation is the structural manifestation of the charge-disproportionated state and therefore produce distinct local crystal fields at the two cobalt sites \cite{ChargeDiso}.

\begin{figure}[!htbp]
\centering
\includegraphics[width=\linewidth]{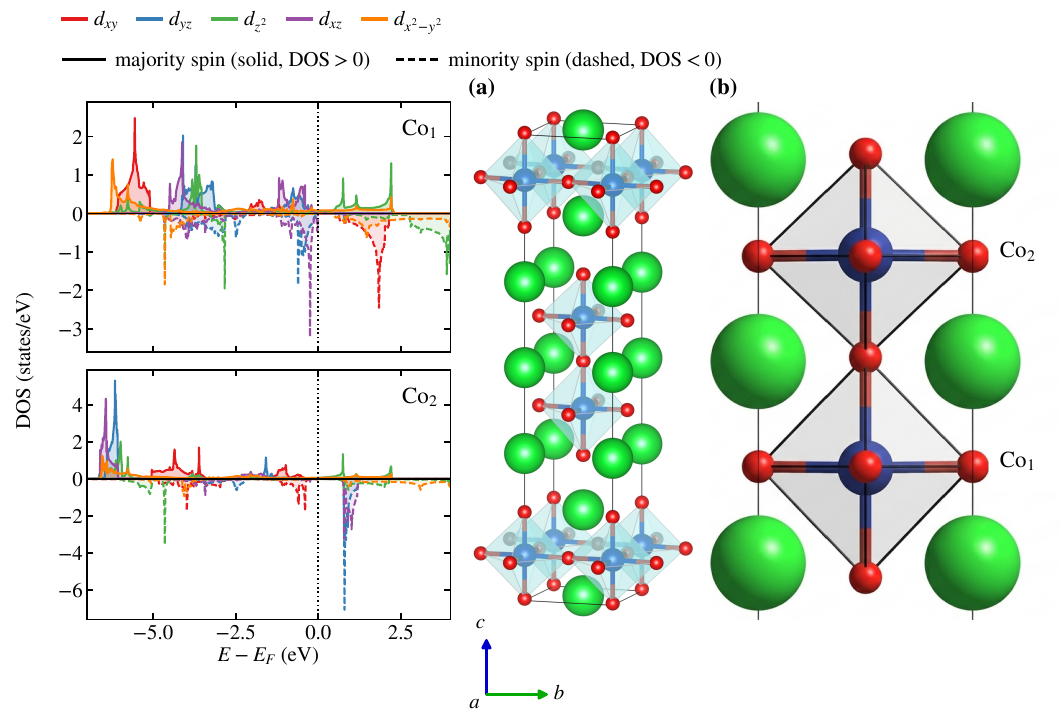}
\caption{\textbf{Charge disproportionation and orbital reconstruction in polar \compound.} Left, scalar-relativistic Co $3d$ projected densities of states for Co1 and Co2. Positive and negative traces denote majority- and minority-spin channels; the dotted line marks $E_F$. Co1 has predominantly occupied minority-spin $d_{xz},d_{yz}$ weight and higher-lying $d_{xy}$ weight, whereas Co2 retains the opposite qualitative hierarchy. Right, \textbf{a}, double-layered Ruddlesden--Popper structure with Sr (green), Co (blue) and O (red); \textbf{b}, local coordination environments for the polar state (POL). The displaced Co1 site has asymmetric apical bonding.}
\label{fig:structure_pdos}
\end{figure}

Figure~\ref{fig:structure_pdos} shows the resulting orbital reconstruction. At Co1, the minority-spin $d_{xz},d_{yz}$ weight lies predominantly below and near $E_F$, whereas $d_{xy}$ shifts upwards. Co2 retains the opposite qualitative hierarchy, with lower-lying $d_{xy}$ weight and partially occupied $d_{xz},d_{yz}$ bands. Charge disproportionation thus changes the relative arrangement of the orbitals participating in the low-energy electronic structure. Thus, the polar state changes the local orbital hierarchy rather than simply shifting all cobalt states together.

The distinction between a band and a discrete atomic level is important here. The projected spectra describe broadened, hybridized states: appreciable weight below the Fermi energy does not establish exact integer occupation of an isolated orbital. We therefore use the level diagrams below to identify the orbital channels allowed by SOC, while the projected spectra and calculated moments describe the actual electronic response.

\FloatBarrier

\subsection{Giant anisotropy accompanies a directional orbital response}
We define the anisotropy from self-consistent SOC total energies as
\begin{equation}
K=E_{\mathrm{tot}}^{[110]}-E_{\mathrm{tot}}^{[001]},
\label{eq:K_energy}
\end{equation}
so that $K<0$ favors [110]. The cell contains four Co atoms, or two formula units. The polar state has $K=-234.0$ meV per cell, equivalent to $-58.5$ meV per Co. The [100] orientation is also stabilized relative to [001], by 193.5 meV per cell, and [110] lies a further 40.5 meV below [100] (Table~\ref{tab:mae}). Thus, the calculated energy landscape strongly favors in-plane magnetization and resolves an additional anisotropy within the plane; [110] is the lowest-energy direction among those examined.

\begin{table}[!htbp]
\centering
\caption{\textbf{Magnetic orientation energies and spin moments in POL.} Energies are relative to [001] and refer to the four-Co cell. $m_s$ denotes the local spin magnetic moment.}
\label{tab:mae}
\begin{tabular}{lrrr}\toprule
Direction & $\Delta E$ (meV/cell) & $m_{s,\mathrm{Co1}}$ ($\mu_B$) & $m_{s,\mathrm{Co2}}$ ($\mu_B$)\\\midrule
$[001]$ & 0.0 & 1.445 & 2.220\\
$[100]$ & $-193.5$ & 1.605 & 2.242\\
$[110]$ & $-234.0$ & 1.632 & 2.253\\\bottomrule
\end{tabular}
\end{table}

The orbital response is particularly strong at Co1. Its orbital-angular-momentum magnitude increases from $0.013\hbar$ for [001] to $1.321\hbar$ for [110]. Co2 also develops a larger in-plane orbital moment, changing from $0.219\hbar$ to $0.717\hbar$ (Table~\ref{tab:moments}). The larger response at the displaced site connects the magnetic orientation energy to the site-selective crystal-field reconstruction. The associated changes in spin moments are smaller, although they reveal a self-consistent electronic response to magnetization rotation.

Orbital unquenching requires coherence between orbital components. Within the $d_{xz},d_{yz}$ sector, complex combinations $(d_{xz}\pm\ii d_{yz})/\sqrt{2}$ can carry angular momentum along $z$. For [110], the relevant operator is $L_{110}=(L_x+L_y)/\sqrt{2}$, which also couples this sector to other Co $d$ orbitals. The reconstructed $t_{2g}$ hierarchy therefore supplies an orbitally responsive environment, while the full $d$ manifold determines the direction of the recovered orbital moment. Supplementary Notes develop this connection explicitly.

\begin{table}[!htbp]
\centering
\caption{\textbf{Directional orbital response and local SOC parameters.} $L_{\hat{\bm n}}$ in this table denotes $|\langle\mathbf L\rangle_{\hat{\bm n}}|$, in units of $\hbar$. The values of $\xi_i$ are the effective moment-product parameters used in the local numerical analysis (Methods); each sublattice contains two Co sites.}
\label{tab:moments}
\begin{tabular}{lrrrrr}\toprule
Site & $L_{110}/\hbar$ & $L_{001}/\hbar$ & $m_{s,110}/\mu_B$ & $m_{s,001}/\mu_B$ & $\xi_i$ (meV)\\\midrule
Co1 & 1.321 & 0.013 & 1.632 & 1.445 & 42.5\\
Co2 & 0.717 & 0.219 & 2.253 & 2.220 & 32.8\\\bottomrule
\end{tabular}
\end{table}
\FloatBarrier

\subsection{Local SOC analysis connects orbital unquenching to the energy scale}
We analyze the SOC response using the orbital-moment contribution $K^{(1)}$ and the diagonal local second-order susceptibility contribution $K^{(2)}$, retaining the notation of the local analysis. TDirect evaluation of the first-order Green-function trace gives
$K^{(1)}=-255.7$ meV per four-Co cell. Co1 contributes $-181.7$ meV and Co2 contributes $-74.0$ meV, so approximately 71\% of this estimate originates from the displaced Co1 sublattice. The same site that shows the strongest orbital recovery therefore supplies the dominant local SOC energy scale.

\begin{figure}[!htbp]
\centering
\includegraphics[width=0.86\linewidth]{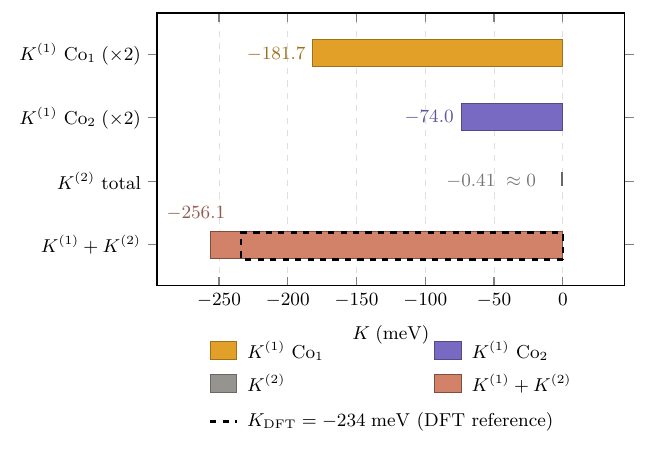}
\caption{\textbf{Local orbital response captures the giant anisotropy scale.} The $K^{(1)}$ contributions are obtained from the site-resolved
first-order Green-function trace, summed over the two Co1
and two Co2 sites, respectively. $K^{(2)}$ is the diagonal local susceptibility estimate. Their sum is compared with $K_{\mathrm{DFT}}=-234.0$ meV per four-Co cell. Co1 dominates the orbital contribution. The combined estimate is $-256.11$ meV, leaving a residual of $+22.11$ meV. The two numerical channels constitute an approximate local analysis rather than an exact partition of the self-consistent DFT energy.}
\label{fig:mae_decomp}
\end{figure}

The conventional local susceptibility analysis gives $K^{(2)}=-0.41$ meV per cell: $-0.47$ meV from Co1 and $+0.06$ meV from Co2. Its magnitude is less than 0.2\% of the self-consistent anisotropy. Within this approximation, virtual-transition contributions are too small to explain the large orientation energy. The dominant numerical scale instead follows the directional orbital response (Fig.~\ref{fig:mae_decomp}).

The combined estimate is $-256.11$ meV per cell, about 9.4\% larger in magnitude than the DFT result. This agreement supports the orbital interpretation at the level of sign, scale and site dependence. The numerical $K^{(1)}$ is obtained from the first-order
Green-function trace. Within the adopted collinear
moment-factorization approximation, the moment-product
expression represents the same contribution, rather than
an additional energy term. The distinction between this estimate, strict perturbative coefficients and total-energy differences is specified in Methods and derived in Supplementary Notes.

\FloatBarrier

\begin{figure}[!htbp]
\centering
\includegraphics[width=0.9\linewidth]{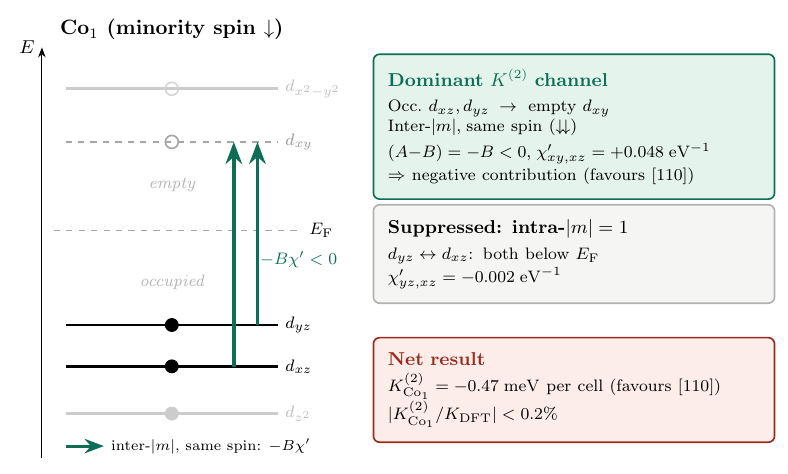}\par\vspace{0.5em}
\includegraphics[width=0.9\linewidth]{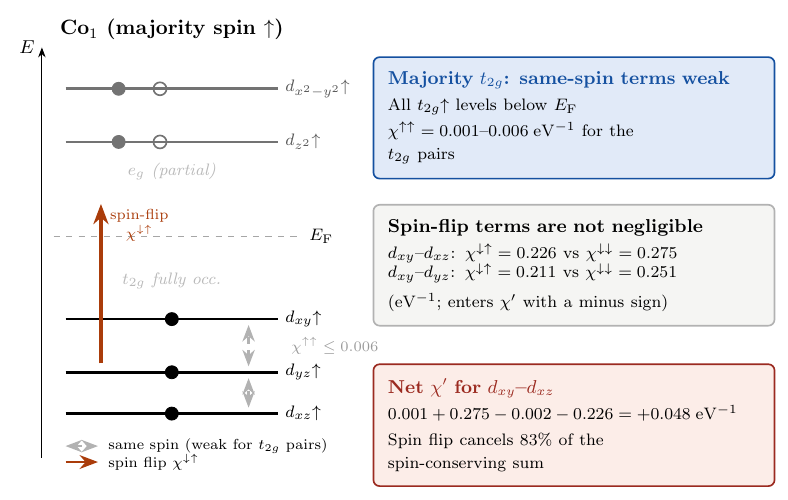}
\caption{\textbf{Local orbital channels at Co1.} Top, minority-spin level schematic for the reconstructed $t_{2g}$ hierarchy. Bottom, idealized occupied majority-spin $t_{2g}$ sector. Filled and open circles illustrate occupied and empty levels; the actual projected states are hybridized bands. The panels identify candidate SOC channels. The reported Co1 sublattice contribution is $K^{(2)}=-0.47$ meV per four-Co cell.}
\label{fig:co1_selection}
\end{figure}

\begin{figure}[!htbp]
\centering
\includegraphics[width=0.9\linewidth]{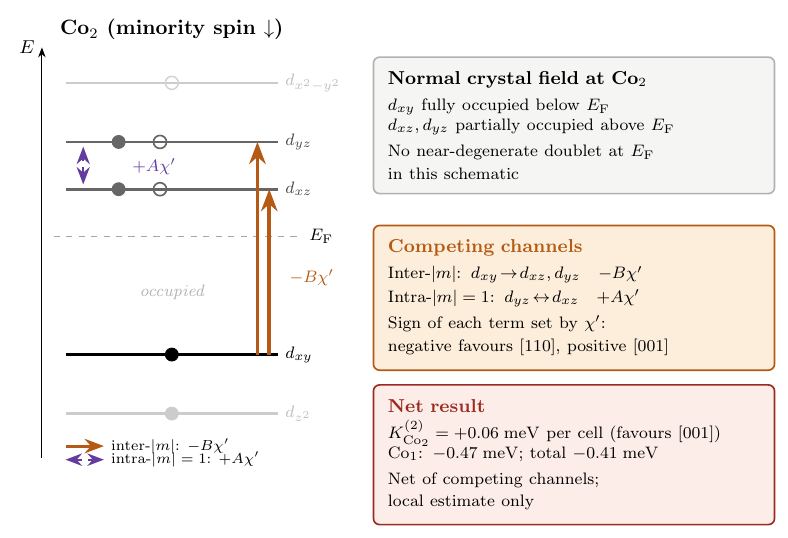}\par\vspace{0.5em}
\includegraphics[width=0.9\linewidth]{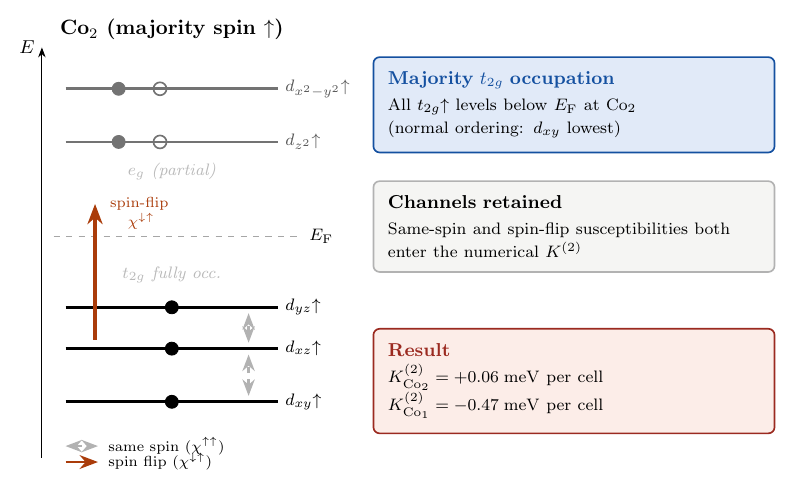}
\caption{\textbf{Local orbital channels at Co2.} Top, minority-spin hierarchy with lower-lying $d_{xy}$ and partially occupied $d_{xz},d_{yz}$ weight. Bottom, idealized majority-spin $t_{2g}$ occupation. The diagrams are qualitative level representations; the partially occupied bands retain both occupied and unoccupied spectral weight. Both spin-conserving and spin-flip terms are retained in the susceptibility analysis. Competing channels yield $K^{(2)}=+0.06$ meV for the Co2 sublattice per four-Co cell.}
\label{fig:co2_selection}
\end{figure}

The orbital channels clarify the small second-order correction. At Co1, transitions between the predominantly occupied minority-spin $d_{xz},d_{yz}$ sector and higher-lying $d_{xy}$ states favor in-plane anisotropy in the local analysis (Fig.~\ref{fig:co1_selection}). At Co2, the reversed hierarchy permits competing channels that give a small positive net contribution (Fig.~\ref{fig:co2_selection}). Spin-conserving and spin-flip susceptibilities enter with opposite signs, and their cancellation contributes to the small net result. Supplementary Notes reports the numerical susceptibilities and their angular weighting.

\FloatBarrier
\FloatBarrier

\subsection{The centrosymmetric reference links the enhancement to orbital reconstruction}
The centrosymmetric reference (SYM) contains four equivalent Co sites and lacks the Co1--Co2 electronic differentiation. Each Co retains a small orbital-angular-momentum magnitude, $0.0523\hbar$ for [110] and $0.0460\hbar$ for [001], with a spin moment of approximately $2.548\mu_B$ for both orientations. Its directional orbital difference is approximately 208 times smaller than that of Co1 in POL.

The self-consistent anisotropy is correspondingly reduced to $-3.05$ meV per four-Co cell, compared with $-234.0$ meV in POL. The local estimates are also much smaller: $K^{(1)}=-7.02$ meV and $K^{(2)}=+4.46$ meV, with partial cancellation. The full comparison is given in Supplementary Note \cite{Supplemental}

Both structures contain the same elements and layered framework, yet their anisotropies differ by approximately a factor of 77. This contrast supports a mechanism in which charge disproportionation and its associated displacement activate the directional orbital response. Because the charge redistribution and structural displacement change together, the comparison identifies their combined role in the enhancement.

\section{Discussion}
The central result is that charge disproportionation can provide a route to large magnetocrystalline anisotropy by reconstructing the orbital environment of a correlated transition-metal oxide. In \compound, the charge-disproportionated polar state combines a large magnetic orientation energy with a pronounced recovery of orbital angular momentum at the displaced cobalt site. The centrosymmetric reference retains nearly quenched orbital moments and much weaker anisotropy. Taken together, these observations connect the enhancement to the electronic and structural differentiation of the Co sublattices.

The microscopic sequence begins with inequivalent local bonding. Charge disproportionation and the accompanying polar displacement modify the Co--O environment, invert the local $t_{2g}$ hierarchy at Co1 and redistribute orbital weight near the Fermi level. This creates low-energy orbital components that can mix under SOC. The mixing depends on magnetization direction: in-plane magnetization supports a much larger orbital moment at Co1 than out-of-plane magnetization. The resulting directional orbital response supplies a large SOC energy scale, as reflected in the dominance of Co1 in the local moment analysis. This sequence explains why the structural reconstruction can amplify anisotropy even though the constituent magnetic ions are $3d$ elements.
The present results connect with the experimental observations of polar metallic Sr$_3$Co$_2$O$_7$ by Zhou et al. \cite{zhou_geometry-driven_2025}. Atomic-resolution microscopy revealed strongly unequal cobalt displacements in adjacent layers, establishing the structural differentiation central to the orbital reconstruction considered here. Co-edge XMCD further identified an appreciable orbital contribution, approximately 20\% of the total magnetic moment under the reported measurement conditions. These observations support the relevance of inequivalent local coordination and active orbital degrees of freedom. Our calculations resolve a strongly site-dependent directional response: the displaced Co1 site develops a substantially larger orbital moment for in-plane magnetization than for magnetization along the stacking direction and dominates the local orbital contribution to the calculated anisotropy.

This site dependence is particularly relevant to the interpretation of magnetic switching in \cite{zhou_geometry-driven_2025}, where different anisotropies in the two Co layers were invoked to explain sequential layer reversal and the associated anomalous Hall hysteresis. The unequal orbital responses found here suggest a microscopic basis for such a contrast: the polar electronic reconstruction makes the SOC energy strongly dependent on both the cobalt environment and magnetic orientation. The local contributions evaluated in the present work, however, are approximate SOC estimates for the ferromagnetic configuration and should not be identified directly with independent layer reversal barriers.

The comparison also requires distinguishing the magnetic configurations. Ref. \cite{zhou_geometry-driven_2025} interprets the experimental state as canted A-type antiferromagnetic order with a Néel vector along the c axis and a low-temperature in-plane net moment, whereas the present anisotropy compares orientations of a ferromagnetic configuration. The calculated in-plane preference therefore provides a qualitative connection to the directionality of the observed net magnetization, but does not determine the antiferromagnetic easy axis or quantitatively reproduce the experimental anisotropy. Extending the analysis to the canted antiferromagnetic state and measuring the angular dependence of the orbital moment would directly test whether the same orbital reconstruction governs the experimental magnetic response.

The connection to first-order SOC physics is the survival of an active orbital manifold. In an appropriately occupied degenerate subspace, SOC can split levels at first order; crystal-field splitting, hybridization and band filling determine whether that linear contribution survives in the total energy \cite{huang_magnetic_2022}. The present local analysis finds a large orbital contribution and a small conventional second-order correction. Its quantitative interpretation is a mechanism based on orbital unquenching; assigning a strict perturbative order additionally requires resolving the SOC dependence of the occupied electronic state. The detailed derivations in the Supplementary Information separate these questions while retaining the local analysis used here.

Layered perovskites offer a promising setting in which to explore this design principle. Their slab geometry and local coordination environments allow composition, epitaxial strain and structural displacements to tune orbital hierarchies \cite{gradauskaite_layered_2026}. The broader potential of orbital-level engineering is
illustrated by predictions of giant magnetic anisotropy
in strained monolayer LiCoTe, where tuning the
$d_{xz/yz}$ and $d_{x^2-y^2}$ orbital multiplet controls the anisotropy \cite{Yao2026}.
The present cobaltate provides a complementary setting
in which charge disproportionation and polar displacement
reconstruct inequivalent local orbital environments
within a layered metal The POL--SYM contrast also shows that layering alone is insufficient: the useful ingredient is the electronically inequivalent, orbitally responsive state that the layered structure can support. Magnetic polar metals extend this opportunity to systems in which structural asymmetry coexists with itinerant carriers. Their potential lies in coupling polarity-generating electronic reconstruction to the same orbital degrees of freedom that govern magnetic orientation.

For the search for rare-earth-free permanent magnets, the implication is a microscopic route to the anisotropy needed for magnetic stability. Candidate materials should combine charge disproportionation, favorable orbital occupation and sufficiently weak quenching to support a large directional SOC response. Strain and chemical substitution provide possible ways to tune these ingredients in related layered oxides. Practical permanent-magnet performance would additionally require adequate magnetization, magnetic ordering temperature and coercivity; these quantities are not established by the present MAE calculations. The result therefore motivates exploration of layered perovskites and polar metals as platforms for anisotropy design, with \compound\ illustrating how charge-driven orbital reconstruction can generate the relevant energy scale.

\section{Methods}
\subsection{Electronic-structure calculations}
Calculations were performed using the Vienna \textit{ab initio} Simulation Package (VASP) \cite{kresse_efficiency_1996,kresse_efficient_1996} with projector augmented-wave potentials\cite{blochl_projector_1994,kresse_ultrasoft_1999} and the Perdew--Burke-Ernzerhof generalized-gradient functional.\cite{perdew_generalized_1996} Co $3d$ correlations were treated using the Liechtenstein DFT + formulation $U$ \cite{Lichtenstein1995} with $U=5.0$ eV and $J=0.9$ eV. The plane-wave cutoff was 550 eV and the Brillouin zone was sampled on a $20\times20\times4$ Monkhorst--Pack mesh. Lattice constants were fixed at $a=b=3.868$ \AA\ and $c=19.80$ \AA. Internal coordinates were relaxed to forces below 1 meV.\AA$^{-1}$.

The reported anisotropies were obtained from fully self-consistent SOC total energies for the ferromagnetic configurations with magnetization along [001], [100] and [110]. Local spin and orbital moments were extracted from the corresponding SOC calculations. A separate scalar-relativistic calculation supplied the orbital-resolved spectra for the unperturbed Green function. POL and SYM were treated using the same functional parameters, interaction parameters, cutoff, and sampling. All cell energies are normalized to four Co atoms. 

\subsection{Green-function description of the SOC response}

We write the effective one-electron Hamiltonian as
$\Htot=\Hzero+\Vso$, where $\Hzero$ is the scalar-relativistic
reference Hamiltonian and
\begin{equation}
\Vso(\hat{\bm n})
=
\sum_i \xi_i\,\mathbf l_i\cdot\mathbf s(\hat{\bm n}).
\end{equation}
Here $\mathbf l_i=\mathbf L_i/\hbar$ and
$\mathbf s=\mathbf S/\hbar$ are dimensionless angular-momentum
operators, and $\hat{\bm n}$ specifies the magnetization direction
relative to the crystal axes. The coefficients $\xi_i$ parameterize
the SOC operator.

The retarded reference and SOC Green functions are
\begin{equation}
\Gz(E)=[E+\ii\eta-\Hzero]^{-1},
\qquad
\G(E)=[E+\ii\eta-\Htot]^{-1},
\qquad \eta>0,
\label{eq:G_defs}
\end{equation}
with spectral matrices
\begin{equation}
\mathbf A_0(E)
=
-\frac{\Gz(E)-\Gz^\dagger(E)}{2\pi\ii},
\qquad
\mathbf A(E)
=
-\frac{\G(E)-\G^\dagger(E)}{2\pi\ii}.
\end{equation}
Occupied-state integrals define the reference density matrix,
\begin{equation}
\boldsymbol{\rho}_0
=
\int_{-\infty}^{E_F}\mathbf A_0(E)\,dE,
\end{equation}
and the corresponding orbital expectation values. For example,
\begin{equation}
\langle L_{\hat{\bm n}}\rangle_{0,i}
=
\int_{-\infty}^{E_F}
\Tr_i\!\left[
\mathbf A_{0,i}(E)L_{\hat{\bm n}}
\right]\,dE.
\end{equation}
Replacing $\mathbf A_0$ by $\mathbf A$ gives the orbital expectation
in the SOC state. These reference-state and SOC-state quantities
are distinguished throughout the analysis.

Introducing $\Htot(\lambda)=\Hzero+\lambda\Vso$ and expanding
the Green function in $\lambda$ separates the contributions
containing one and two insertions of the SOC operator.
Within the fixed-reference Fermi-sea expansion,
\begin{align}
E^{(1)}(\hat{\bm n})
&=
-\frac{1}{\pi}\ImPart
\int_{-\infty}^{E_F}
\Tr\!\left[
\Gz(E)\Vso(\hat{\bm n})
\right]\,dE,
\label{eq:E1_GF}
\\
E^{(2)}(\hat{\bm n})
&=
-\frac{1}{2\pi}\ImPart
\int_{-\infty}^{E_F}
\Tr\!\left[
\Gz(E)\Vso(\hat{\bm n})
\Gz(E)\Vso(\hat{\bm n})
\right]\,dE.
\label{eq:E2_GF}
\end{align}
The different prefactors follow from coupling-constant integration.
The anisotropy convention is
\begin{equation}
K^{(r)}
=
E^{(r)}([110])-E^{(r)}([001]),
\end{equation}
so that negative values favour [110].


\subsection{First-order orbital contribution and second-order
susceptibility analysis}

The reported first-order anisotropy was evaluated directly from
the Green-function trace,
\begin{equation}
K^{(1)}
=
-\frac{1}{\pi}\ImPart
\int_{-\infty}^{E_F}
\Tr\!\left\{
\Gz(E)
\left[
\Vso([110])-\Vso([001])
\right]
\right\}\,dE.
\label{eq:K1_formula}
\end{equation}
For an on-site SOC operator, the trace can be resolved into
atomic contributions,
\begin{equation}
E^{(1)}(\hat{\bm n})
=
\sum_i \xi_i
\Tr_i\!\left[
\boldsymbol{\rho}_{0,i}\,
\mathbf l_i\cdot\mathbf s(\hat{\bm n})
\right].
\label{eq:E1_density}
\end{equation}

For a spin-diagonal collinear reference, the first-order
expectation reduces to the longitudinal spin component:
\begin{equation}
E^{(1)}(\hat{\bm n})
=
\frac{1}{2}\sum_i\xi_i
\Tr_{\mathrm{orb}}\!\left[
\left(
\boldsymbol{\rho}_{0,i}^{\uparrow}
-
\boldsymbol{\rho}_{0,i}^{\downarrow}
\right)
l_{i,\hat{\bm n}}
\right].
\label{eq:E1_collinear}
\end{equation}
Equivalently,
\begin{equation}
E^{(1)}(\hat{\bm n})
=
\sum_i\frac{\xi_i}{2\hbar}
\left[
L_{0,i,\hat{\bm n}}^{\uparrow}
-
L_{0,i,\hat{\bm n}}^{\downarrow}
\right],
\label{eq:E1_orbital}
\end{equation}
where
$L_{0,i,\hat{\bm n}}^{\sigma}
=
\Tr_{\mathrm{orb}}
[
\boldsymbol{\rho}_{0,i}^{\sigma}L_{i,\hat{\bm n}}
]$
is a signed, spin-resolved orbital-angular-momentum expectation.
Equations~\eqref{eq:E1_density}--\eqref{eq:E1_orbital}
express the same first-order contribution and connect the
Green-function trace to its orbital interpretation.

Direct evaluation gives $K^{(1)}=-255.7$ meV per four-Co cell
for POL, with contributions of $-181.7$ and $-74.0$ meV
from the Co1 and Co2 sublattices, respectively.
The self-consistent SOC moments characterize the directional
orbital response at physical SOC strength and are reported
alongside the perturbative analysis.

For the second-order contribution, we retain on-site terms
and use the orbital-diagonal approximation to the reference
Green function.\cite{Li2015}
The orbital-pair susceptibilities are
\begin{equation}
\chi_{i;mm'}^{\sigma\sigma'}
=
\frac{1}{\pi}\ImPart
\int_{-\infty}^{E_F}
g_{im}^{\sigma}(E)
g_{im'}^{\sigma'}(E)\,dE,
\label{eq:chi_def}
\end{equation}
where $g_{im}^{\sigma}$ denotes a diagonal element of the
local scalar-relativistic Green function.
The retained second-order anisotropy is
\begin{equation}
K^{(2)}
=
\sum_i\frac{\xi_i^2}{8}
\sum_{mm'}
(A_{mm'}-B_{mm'})
\chi'_{i;mm'},
\label{eq:K2_AB}
\end{equation}
with
\begin{equation}
A_{mm'}
=
|\langle m|l_z|m'\rangle|^2,
\qquad
B_{mm'}
=
\frac{1}{2}
|\langle m|l_x+l_y|m'\rangle|^2,
\end{equation}
and
\begin{equation}
\chi'_{i;mm'}
=
\chi_{i;mm'}^{\uparrow\uparrow}
+
\chi_{i;mm'}^{\downarrow\downarrow}
-
\chi_{i;mm'}^{\uparrow\downarrow}
-
\chi_{i;mm'}^{\downarrow\uparrow}.
\end{equation}
Both spin-conserving and spin-flip channels are therefore
included. 
The second-order approximation omits intersite and
orbital off-diagonal propagation. Its orbital-diagonal
restriction is not imposed on the formal first-order
expression in equation~\eqref{eq:E1_density}.
For POL, the retained second-order contribution is
$-0.41$ meV per four-Co cell. The sum
$K^{(1)}+K^{(2)}=-256.11$ meV
is compared with the self-consistent DFT anisotropy
of $-234.0$ meV. This comparison assesses the retained
perturbative contributions; it is not an exact decomposition
of the self-consistent DFT total-energy difference. 

\section{Data availability}
The total-energy differences, local moments and susceptibility values discussed in this work are provided in the article and Supplementary Information. The underlying susceptibility data are available from the corresponding author upon reasonable request.

\section*{Code availability}
The source code to reproduce the Green function results is available in the accompanying GitHub repository \href{https://www.github.com/hsmsbri/MAEGreenFunction}{https://www.github.com/hsmsbri/MAEGreenFunction/}
\section{Acknowledgements}
This research used resources of the National Energy Research Scientific Computing Center, a DOE Office of Science User Facility, using NERSC award BES-ERCAP0030590. The authors gratefully acknowledge the support and computing resources from the Toubkal Supercomputer (\href{https://toubkal.um6p.ma}{https://toubkal.um6p.ma})\cite{kissami2025toubkal}.

\section{Author contributions}
J.-X.Y. and J.Z. conceived the research project. H.S., H.-F.H. and J.-X.Y. performed the DFT calculations. H.S. performed the Green-function analysis and prepared the figures. J.Z. contributed to the theoretical analysis and interpretation of the results. J.-X.Y. and J.Z. supervised the research and provided critical feedback. All authors contributed to writing and revising the manuscript and approved the final version.

\section{Competing interests}
The authors declare no competing interests.

\begingroup\small\setstretch{1.0}\setlength{\bibsep}{2pt}\setlength{\parskip}{0pt}

\endgroup
\clearpage 
%
%
%
%
%


\setstretch{1.10}
\setlength{\parindent}{0pt}
\setlength{\parskip}{0.45em}
\setlength{\emergencystretch}{2em}

\setcounter{secnumdepth}{0}
\setcounter{tocdepth}{1}

\titleformat{\section}
  {\large\bfseries}{}{0pt}{}

\titleformat{\subsection}
  {\normalsize\bfseries}{}{0pt}{}

\titlespacing*{\section}{0pt}{1.3em}{0.5em}
\titlespacing*{\subsection}{0pt}{0.8em}{0.2em}

\captionsetup{
  font=small,
  labelfont=bf,
  labelsep=period,
  justification=justified,
  singlelinecheck=false
}

\setlength{\textfloatsep}{14pt plus 3pt minus 2pt}
\setlength{\intextsep}{12pt plus 2pt minus 2pt}


\renewcommand{\theequation}{S\arabic{equation}}
\renewcommand{\thetable}{S\arabic{table}}
\renewcommand{\thefigure}{S\arabic{figure}}


\begin{center}
  {\Large\bfseries Supplementary Information\par}

  \vspace{0.5em}

  {\large\bfseries \papertitle\par}

  \vspace{0.5em}

  Houssam Sabri, Hong-Fei Huang, Jie-Xiang Yu and Jiadong Zang
\end{center}

\section{Green-function formulation}

We write the effective one-electron Hamiltonian as
\begin{equation}
\Htot=\Hzero+\Vso,
\qquad
\Vso=\sum_i\xi_i\,\mathbf l_i\cdot\mathbf s,
\label{eq:H_split}
\end{equation}
where $\Hzero$ is the scalar-relativistic Hamiltonian,
$\xi_i$ is the local SOC parameter, and
$\mathbf l=\mathbf L/\hbar$ and
$\mathbf s=\mathbf S/\hbar$ are dimensionless angular momenta.
The retarded Green functions are
\begin{equation}
\G(E)=[E+\ii\eta-\Htot]^{-1},
\qquad
\Gz(E)=[E+\ii\eta-\Hzero]^{-1}.
\label{eq:G_defs}
\end{equation}
Here $\eta>0$ is the spectral broadening.
The spectral matrix is
$\mathbf A(E)=-(\G-\G^\dagger)/(2\pi\ii)$,
and the band energy at zero temperature is
\begin{equation}
E_{\mathrm{band}}
=
\int_{-\infty}^{E_F}
E\,\Tr\mathbf A(E)\,dE.
\end{equation}

Introducing $\Htot(\lambda)=\Hzero+\lambda\Vso$,
the SOC-induced band-energy change at fixed electron number is
\begin{equation}
\Delta E_{\mathrm{band}}
=
-\frac{1}{\pi}\ImPart
\int_0^1d\lambda
\int_{-\infty}^{E_F(\lambda)}
\Tr[\G_\lambda(E)\Vso]\,dE,
\label{eq:cc_int}
\end{equation}
where $E_F(\lambda)$ maintains the electron number.
The Dyson expansion is
\begin{equation}
\G_\lambda
=
\Gz+\lambda\Gz\Vso\Gz
+\lambda^2\Gz\Vso\Gz\Vso\Gz+\cdots.
\label{eq:Dyson_series}
\end{equation}

For the fixed-reference Fermi-sea expansion used in the local
analysis, integration over $\lambda$ gives
\begin{align}
E^{(1)}(\hat{\bm n})
&=
-\frac{1}{\pi}\ImPart
\int_{-\infty}^{E_F}
\Tr[\Gz(E)\Vso(\hat{\bm n})]\,dE,
\label{eq:E1_GF}
\\
E^{(2)}(\hat{\bm n})
&=
-\frac{1}{2\pi}\ImPart
\int_{-\infty}^{E_F}
\Tr[
\Gz(E)\Vso(\hat{\bm n})
\Gz(E)\Vso(\hat{\bm n})
]\,dE.
\label{eq:E2_GF}
\end{align}

Here $E_F$ is the scalar-relativistic reference Fermi level;
changes of occupation must be treated consistently if they
contribute. In a regular quadratic regime with vanishing
first-order coefficient, the endpoint functional
$E_r=-\ImPart\int^{E_F}\Tr(\G\Vso)dE/(2\pi)$
reproduces the second-order energy through that order.
The first-order coefficient in equation~\eqref{eq:E1_GF}
has no half-prefactor.

We use the anisotropy convention
\begin{equation}
K=E^{[110]}-E^{[001]},
\qquad
K^{(r)}
=
E^{(r)}([110])-E^{(r)}([001]),
\label{eq:K_def}
\end{equation}
so negative $K$ favors the in-plane [110] direction.

\section{First-order contribution:
\texorpdfstring{$K^{(1)}$}{K1}}

Resolving the first-order trace into local orbital and spin
subspaces gives
\begin{equation}
E^{(1)}(\hat{\bm n})
=
-\frac{1}{\pi}\sum_i\xi_i\ImPart
\int_{-\infty}^{E_F}
\Tr_i[
\mathbf G_{0,i}(E)\,
\mathbf l_i\cdot\mathbf s(\hat{\bm n})
]\,dE.
\label{eq:E1_site}
\end{equation}

The Green-function expression for the first-order anisotropy
is therefore
\begin{equation}
\boxed{
K^{(1)}
=
-\frac{1}{\pi}\ImPart
\int_{-\infty}^{E_F}
\Tr\!\left\{
\Gz(E)
[\Vso([110])-\Vso([001])]
\right\}\,dE.
}
\label{eq:K1_GF}
\end{equation}

The reported POL value $K^{(1)}=-255.7$ meV per four-Co cell
is obtained by direct evaluation of equation~\eqref{eq:K1_GF}.
Its collinear orbital representation follows from the
reference-state density matrix
\begin{equation}
\boldsymbol{\rho}_{0,i}
=
\int_{-\infty}^{E_F}\mathbf A_{0,i}(E)\,dE,
\qquad
\mathbf A_0(E)
=
-\frac{\Gz(E)-\Gz^\dagger(E)}{2\pi\ii}.
\label{eq:rho0}
\end{equation}

For an on-site SOC operator,
equation~\eqref{eq:E1_site} is equivalently
\begin{equation}
E^{(1)}(\hat{\bm n})
=
\sum_i\xi_i\Tr_i\!\left[
\boldsymbol{\rho}_{0,i}\,
\mathbf l_i\cdot\mathbf s(\hat{\bm n})
\right].
\label{eq:E1_density}
\end{equation}

For a spin-diagonal collinear reference, the off-diagonal spin
blocks do not contribute to this first-order expectation.
In the spin basis aligned with $\hat{\bm n}$,
\begin{align}
E^{(1)}(\hat{\bm n})
&=
\frac12\sum_i\xi_i
\Tr_{\mathrm{orb}}\!\left[
(\boldsymbol{\rho}_{0,i}^{\uparrow}
-\boldsymbol{\rho}_{0,i}^{\downarrow})
l_{i,\hat{\bm n}}
\right]
\label{eq:E1_collinear}
\\
&=
\sum_i\frac{\xi_i}{2\hbar}
\left(
L_{0,i,\hat{\bm n}}^{\uparrow}
-
L_{0,i,\hat{\bm n}}^{\downarrow}
\right),
\label{eq:E1_spinweighted}
\end{align}
where
$L_{0,i,\hat{\bm n}}^{\sigma}
=
\Tr_{\mathrm{orb}}[
\boldsymbol{\rho}_{0,i}^{\sigma}L_{i,\hat{\bm n}}
]$
is a signed spin-resolved orbital-angular-momentum expectation.
These expressions are representations of the same first-order
trace. In particular, the spin weighting involves the difference
between the two spin-resolved orbital moments.

The orbital expectation in the SOC state is obtained separately as
\begin{equation}
\langle L_{\hat{\bm n}}\rangle_i
=
-\frac{1}{\pi}\ImPart
\int_{-\infty}^{E_F}
\Tr_i[\G_i(E)L_{\hat{\bm n}}]\,dE.
\label{eq:L_from_G}
\end{equation}

Using $\Gz$ in this expression gives the reference-state
expectation. In a complex-harmonic basis,
\begin{align}
\rho_{im\sigma}(E)
&=
-\frac{1}{\pi}\ImPart
\mel{i,m,\sigma}{\G(E)}{i,m,\sigma},
\\
\langle L_z\rangle_i
&=
\hbar\sum_{m,\sigma}m
\int_{-\infty}^{E_F}
\rho_{im\sigma}(E)\,dE.
\end{align}

The self-consistent SOC moments characterize the response at
physical SOC strength; they are distinguished from the
reference-state quantities in
equations~\eqref{eq:E1_density}--\eqref{eq:E1_spinweighted}.

\subsection{Why diagonal real-orbital populations are insufficient}

In the real basis
$P=\{\ket{d_{xz}},\ket{d_{yz}}\}$,
the diagonal matrix elements of $L_z$ vanish, whereas its
off-diagonal elements are finite:
\begin{equation}
L_z\big|_P
=
\hbar
\begin{pmatrix}
0&-\ii\\
\ii&0
\end{pmatrix},
\qquad
\ket{\pm}
=
\frac{\ket{d_{xz}}\pm\ii\ket{d_{yz}}}{\sqrt2}.
\label{eq:orbital_split}
\end{equation}

For an exactly degenerate doublet and a fixed dimensionless spin
projection $s_z$, the longitudinal projected SOC is
$\Vso^{(P)}=\xi s_z L_z/\hbar$.
Its eigenstates $\ket{\pm}$ carry opposite angular momenta
$\pm\hbar$ and acquire shifts
$E_\pm^{(1)}=\pm\xi s_z$,
giving a splitting $2|\xi s_z|$ that is linear in SOC.
Unequal occupation of these states produces an orbital moment;
equal occupation cancels it.

Diagonal real-orbital populations alone cannot distinguish these
situations because both complex states contain equal $d_{xz}$
and $d_{yz}$ weights. Writing the local density matrix as
$\varrho_i=\int_{-\infty}^{E_F}\mathbf A_i(E)\,dE$,
the doublet contribution is
$\langle L_z\rangle_{i,P}
=-2\hbar\,\ImPart(\varrho_{i;xz,yz})$,
with a spin sum understood. It therefore depends on the
off-diagonal orbital coherence retained in the Green-function
trace, not simply on the difference between the diagonal
$d_{xz}$ and $d_{yz}$ populations.

For [110], the relevant operator is
$L_{110}=(L_x+L_y)/\sqrt2$,
which couples the reconstructed $t_{2g}$ sector to other
components of the full Co $d$ manifold. The doublet splitting
illustrates the unquenching mechanism; the large in-plane
orbital moment requires this full directional response.

\subsection{Collinear moment representation of the first-order term}

The local moment-product expression describes the same
first-order contribution within the adopted collinear
moment-factorization approximation:
\begin{equation}
\boxed{
K^{(1)}
\simeq
-\sum_i\xi_i^{\mathrm{eff}}
\left[
\frac{L_{i,110}}{\hbar}
\frac{|m_{s,i,110}|}{\mu_B}
-
\frac{L_{i,001}}{\hbar}
\frac{|m_{s,i,001}|}{\mu_B}
\right].
}
\label{eq:K1_formula}
\end{equation}

Here $L_{i,n}=|\langle\mathbf L_i\rangle_n|$ and $m_{s,i,n}$
are the local moments reported for the SOC calculations.
Equation~\eqref{eq:K1_formula} is a moment representation of
$K^{(1)}$, not an additional contribution to be added to the
first-order trace.

The relation to the spin-resolved expression can be stated
explicitly. Define
\begin{equation}
Q_{i,n}
=
\frac{L_{i,n}}{\hbar}
\frac{|m_{s,i,n}|}{\mu_B},
\qquad
c_{i,n}
=
-\frac{E_i^{(1)}(n)}{Q_{i,n}},
\label{eq:moment_mapping}
\end{equation}
for nonzero $Q_{i,n}$.
Equation~\eqref{eq:K1_formula} follows when the directional
coefficients $c_{i,n}$ are represented by a common effective
coefficient $\xi_i^{\mathrm{eff}}$ at each site.
This step incorporates the local spin weighting, moment
normalization and sign convention. The use of moment magnitudes
requires the signed energy contribution to be encoded
consistently in that coefficient; collinearity alone does not
fix its sign or directional dependence.

The reported effective coefficients use the normalization
\begin{equation}
\xi_i^{\mathrm{eff}}
=
\frac{|E_{\mathrm{soc},i}^{[110]}|}
{(L_{i,110}/\hbar)(|m_{s,i,110}|/\mu_B)}.
\label{eq:xi_eff}
\end{equation}

The superscript ``eff'' distinguishes this moment-product
normalization from the coefficient of the one-electron SOC
operator in equation~\eqref{eq:H_split}.
The numerical POL first-order result is assigned from
equation~\eqref{eq:K1_GF};
equation~\eqref{eq:K1_formula} provides its local moment
interpretation. Identifying the endpoint normalization in
equation~\eqref{eq:xi_eff} with the reference-state coefficients
in equation~\eqref{eq:moment_mapping} is part of the
moment-factorization approximation, not a general identity
between an SOC-state expectation and a perturbative coefficient.

\section{Second-order contribution:
\texorpdfstring{$K^{(2)}$}{K2}}

Starting from equation~\eqref{eq:E2_GF}, we retain the on-site
terms and use the orbital-diagonal approximation for this
second-order calculation. This restriction is separate from
the orbital matrix structure required by the first-order trace:
\begin{equation}
[G_{0,i}(E)]_{m\sigma,m'\sigma'}
\simeq
\delta_{mm'}\delta_{\sigma\sigma'}
g_{im}^{\sigma}(E).
\end{equation}

The trace then becomes a sum over orbital pairs:
\begin{equation}
\Tr_i(
\mathbf G_{0,i}v_i
\mathbf G_{0,i}v_i
)
=
\xi_i^2
\sum_{mm',\sigma\sigma'}
g_{im}^{\sigma}g_{im'}^{\sigma'}
\left|
\mel{m\sigma}
{\mathbf l\cdot\mathbf s(\hat{\bm n})}
{m'\sigma'}
\right|^2,
\end{equation}
where $v_i=\xi_i\mathbf l_i\cdot\mathbf s$.

Defining the orbital-pair susceptibility\cite{Li2015}
\begin{equation}
\chi_{i;mm'}^{\sigma\sigma'}(E_F)
=
\frac{1}{\pi}\ImPart
\int_{-\infty}^{E_F}
g_{im}^{\sigma}(E)
g_{im'}^{\sigma'}(E)\,dE,
\label{eq:chi_def}
\end{equation}
we obtain
\begin{equation}
E^{(2)}(\hat{\bm n})
=
-\frac12\sum_i\xi_i^2
\sum_{mm',\sigma\sigma'}
\left|
\mel{m\sigma}
{\mathbf l\cdot\mathbf s(\hat{\bm n})}
{m'\sigma'}
\right|^2
\chi_{i;mm'}^{\sigma\sigma'}.
\label{eq:E2_chi}
\end{equation}

In the spin basis aligned with $\hat{\bm n}$,
\begin{equation}
\mathbf l\cdot\mathbf s
=
\frac12
\begin{pmatrix}
l_{\hat{\bm n}}
&
l_{\mathbf e_1}-\ii l_{\mathbf e_2}
\\
l_{\mathbf e_1}+\ii l_{\mathbf e_2}
&
-l_{\hat{\bm n}}
\end{pmatrix},
\end{equation}
where $(\mathbf e_1,\mathbf e_2,\hat{\bm n})$
is an orthonormal triad.
The diagonal and off-diagonal spin blocks produce
spin-conserving and spin-flip terms.
In a real-orbital basis, taking the difference between
[110] and [001] gives
\begin{equation}
\boxed{
K^{(2)}
=
\sum_i\frac{\xi_i^2}{8}
\sum_{mm'}
(A_{mm'}-B_{mm'})
\chi'_{i;mm'},
}
\label{eq:K2_AB}
\end{equation}
with the susceptibility combination
\begin{equation}
\chi'_{i;mm'}
=
\chi_{i;mm'}^{\uparrow\uparrow}
+
\chi_{i;mm'}^{\downarrow\downarrow}
-
\chi_{i;mm'}^{\uparrow\downarrow}
-
\chi_{i;mm'}^{\downarrow\uparrow}.
\end{equation}

The angular coefficients for the stated orientation comparison
are
\begin{equation}
A_{mm'}
=
|\langle m|l_z|m'\rangle|^2,
\qquad
B_{mm'}
=
\frac12
|\langle m|l_x+l_y|m'\rangle|^2.
\end{equation}

In real-harmonic indexing,
$A_{mm'}=m^2\delta_{m,-m'}$
connects the opposite members of a fixed $|m|$ sector.
The transverse coefficients follow from the usual
angular-momentum selection rules.\cite{Li2015}
The groupings
$|m|=1:\{d_{xz},d_{yz}\}$,
$|m|=2:\{d_{xy},d_{x^2-y^2}\}$
and $m=0:d_{z^2}$
organize these channels without assuming exact orbital
degeneracy. Spin-conserving and spin-flip susceptibilities enter
with opposite signs, and the net contribution is determined
by their angularly weighted sum.

The symbol $K^{(2)}$ denotes the retained local,
orbital-diagonal contribution in the numerical tables.
Intersite and orbital off-diagonal propagation are omitted
at this stage.
The $\xi_i$ entering equations~\eqref{eq:E2_chi}
and~\eqref{eq:K2_AB} are coefficients in the operator convention
of equation~\eqref{eq:H_split}; a moment-product normalization
such as equation~\eqref{eq:xi_eff} requires a consistent
conversion before use in these expressions.


\section{Polar-phase numerical analysis}

The polar phase contains two Co1 and two Co2 sites per cell.
The SOC calculations give the local moments in
Table~\ref{tab:pol_moments}, which also lists their effective
moment-product normalizations.
The displaced Co1 site has the largest directional orbital
response, with
$L_{110}-L_{001}=1.3079\hbar$,
compared with $0.4980\hbar$ at Co2.

\begin{table}[H]
\centering
\caption{
Local SOC-state moments and effective moment-product parameters
in POL. Orbital moments are in $\hbar$, spin magnetic moments
in $\mu_B$, and $\xi^{\mathrm{eff}}$ in meV.
The effective coefficients are defined by
equation~\eqref{eq:xi_eff}.
Each sublattice contains two Co sites.
}
\label{tab:pol_moments}
\begin{tabular}{lrrrrr}
\toprule
Site
&
$L_{110}$
&
$L_{001}$
&
$|m_{s,110}|$
&
$|m_{s,001}|$
&
$\xi^{\mathrm{eff}}$
\\
\midrule
Co1 & 1.3209 & 0.013 & 1.632 & 1.445 & 42.5\\
Co2 & 0.7170 & 0.219 & 2.253 & 2.220 & 32.8\\
\bottomrule
\end{tabular}
\end{table}

Direct evaluation of the site-resolved first-order Green-function
trace in equations~\eqref{eq:E1_site} and~\eqref{eq:K1_GF}
gives $-181.7$ meV from the Co1 sublattice and $-74.0$ meV
from Co2, giving
$K_{\mathrm{POL}}^{(1)}=-255.7$ meV per four-Co cell.
Co1 thus contributes approximately 2.5 times as much as Co2,
reflecting its stronger orbital unquenching.

The moment representation in equation~\eqref{eq:K1_formula},
evaluated with the rounded parameters in
Table~\ref{tab:pol_moments}, approximately reproduces the
reported trace contributions.
This comparison concerns two representations of the first-order
term; it does not supply an additional energy contribution.

The second-order term follows from the scalar-relativistic
orbital-pair susceptibilities in equation~\eqref{eq:chi_def}.
Table~\ref{tab:pol_chi} lists selected nonzero
angular-momentum channels at Co1.
For $d_{xy}\leftrightarrow d_{xz},d_{yz}$,
the positive $\chi'$ combines with $A-B<0$
to favor in-plane anisotropy.
Spin-flip terms substantially reduce the spin-conserving
contribution.
Each susceptibility must be multiplied by its angular factor:
a large diagonal real-orbital susceptibility alone does not
generate a contribution because the corresponding diagonal
angular-momentum matrix element vanishes.

\begin{table}[H]
\centering
\caption{
Selected Co1 orbital-pair susceptibilities in POL,
in eV$^{-1}$, with
$\chi'=
\chi^{\uparrow\uparrow}
+\chi^{\downarrow\downarrow}
-\chi^{\uparrow\downarrow}
-\chi^{\downarrow\uparrow}$.
}
\label{tab:pol_chi}
\begin{tabular}{lrrrrr}
\toprule
Pair
&
$\chi^{\uparrow\uparrow}$
&
$\chi^{\downarrow\downarrow}$
&
$\chi^{\uparrow\downarrow}$
&
$\chi^{\downarrow\uparrow}$
&
$\chi'$
\\
\midrule
$d_{xy}\leftrightarrow d_{xz}$
& 0.001 & 0.275 & 0.002 & 0.226 & $+0.048$\\

$d_{xy}\leftrightarrow d_{yz}$
& 0.002 & 0.251 & 0.003 & 0.211 & $+0.039$\\

$d_{xy}\leftrightarrow d_{x^2-y^2}$
& 0.041 & 0.079 & 0.105 & 0.144 & $-0.129$\\

$d_{yz}\leftrightarrow d_{xz}$
& 0.006 & 0.008 & 0.009 & 0.007 & $-0.002$\\
\bottomrule
\end{tabular}
\end{table}

Summing the retained local second-order channels gives
$K_{\mathrm{Co1}}^{(2)}=-0.47$ meV and
$K_{\mathrm{Co2}}^{(2)}=+0.06$ meV,
including two sites of each type, and hence
$K_{\mathrm{POL}}^{(2)}=-0.41$ meV per cell.
These totals include channels beyond the selected entries
in Table~\ref{tab:pol_chi}.

The sum of the retained first- and second-order contributions
is $-256.11$ meV, compared with
$K_{\mathrm{DFT}}=-234.0$ meV,
leaving a difference of $+22.11$ meV
(9.45\% of $|K_{\mathrm{DFT}}|$).
Within this analysis, the first-order contribution dominates
the retained local second-order term and accounts for the
sign and approximate scale of the DFT anisotropy.
The comparison is not an exact partition of the self-consistent
total energy: higher-order terms, the omitted second-order
matrix elements and self-consistent electronic relaxation
are not separately resolved.

\FloatBarrier

\begingroup
\setstretch{1.0}
\setlength{\parskip}{0.3em}

\section{Centrosymmetric reference and POL--SYM comparison}

In the centrosymmetric SYM reference, the four Co sites are
equivalent and the site-selective orbital reconstruction
associated with charge disproportionation is absent.
The calculated moments are
$L_{110}=0.0523\hbar$,
$L_{001}=0.0460\hbar$
and
$|m_{s,110}|=|m_{s,001}|=2.548\mu_B$.
Thus the orbital-moment anisotropy is only
$0.0063\hbar$ per Co, approximately 208 times smaller than
at POL Co1.
The orbital angular momentum remains largely quenched
in both magnetization directions.

Using the reported moment normalization
$\xi_{\mathrm{SYM}}^{\mathrm{eff}}=108.94$ meV
in equation~\eqref{eq:K1_formula} gives
\begin{equation}
K_{\mathrm{SYM}}^{(1)}
\simeq
-4\xi_{\mathrm{SYM}}^{\mathrm{eff}}
\left[
(0.0523)(2.548)-(0.0460)(2.548)
\right]
\simeq -7.0~\mathrm{meV}.
\end{equation}

The reported first-order value is $-7.02$ meV per cell.
Here $\xi^{\mathrm{eff}}$ retains the moment-product
normalization defined above; the small orbital moments make
this ratio sensitive to rounding and local projection.

The scalar-relativistic Green functions yield the directional
susceptibilities in Table~\ref{tab:sym_chi}.
Their angularly weighted sum gives
$K_{\mathrm{SYM}}^{(2)}=+4.46$ meV per four-Co cell.
This positive contribution favors [001] relative to [110]
and partially cancels the negative first-order term.

\begin{table}[H]
\centering
\caption{
Representative directional susceptibilities $\chi'_{mm'}$
for an equivalent Co site in SYM, in eV$^{-1}$.
}
\label{tab:sym_chi}
\begin{tabular}{lr}
\toprule
Orbital pair & $\chi'$\\
\midrule
$d_{xy}\leftrightarrow d_{yz}$
& $+0.117266$\\

$d_{xy}\leftrightarrow d_{xz}$
& $+0.117266$\\

$d_{xy}\leftrightarrow d_{x^2-y^2}$
& $+0.017023$\\

$d_{yz}\leftrightarrow d_{xz}$
& $-0.135194$\\

$d_{yz}\leftrightarrow d_{z^2}$
& $-0.139708$\\

$d_{xz}\leftrightarrow d_{z^2}$
& $-0.139708$\\

$d_{yz}\leftrightarrow d_{x^2-y^2}$
& $-0.140837$\\

$d_{xz}\leftrightarrow d_{x^2-y^2}$
& $-0.140837$\\
\bottomrule
\end{tabular}
\end{table}

\begin{table}[H]
\centering
\caption{
POL--SYM comparison. Energies are in meV per four-Co cell;
negative values favor [110] relative to [001].
The sums use the displayed rounded contributions.
}
\label{tab:pol_sym}
\begin{tabular}{lrr}
\toprule
Quantity & POL & SYM\\
\midrule
$K^{(1)}$
& $-255.7$ & $-7.02$\\

$K^{(2)}$
& $-0.41$ & $+4.46$\\

$K^{(1)}+K^{(2)}$
& $-256.11$ & $-2.56$\\

$K_{\mathrm{DFT}}$
& $-234.0$ & $-3.05$\\
\bottomrule
\end{tabular}
\end{table}

The DFT anisotropy is about 77 times larger in POL than in SYM.
As summarized in Table~\ref{tab:pol_sym}, the change is
accompanied by a pronounced increase in the first-order
contribution.
In POL, the retained second-order term is negligible on the
scale of $K^{(1)}$; in SYM, it is comparable to $K^{(1)}$
and substantially cancels it.
The comparison therefore supports the orbital-reconstruction
interpretation: charge disproportionation reconstructs
the local cobalt crystal fields, enables a strongly
direction-dependent orbital moment and supplies the large
anisotropy energy scale in the polar phase.

\FloatBarrier


\begingroup
\small

\renewcommand{\refname}{Supplementary reference}

\endgroup
\endgroup

\end{document}